# Cosmic ray modulation of infra-red radiation in the atmosphere

K L Aplin[1] and M Lockwood[2]

1. Physics Department, University of Oxford, Denys Wilkinson Building, Keble Road, Oxford OX1 3RH
2. Space and Atmospheric Electricity Group, Department of Meteorology, University of Reading, Earley Gate, Reading RG6 6BB

Corresponding author: k.aplin1@physics.ox.ac.uk

**Abstract**. Cosmic rays produce molecular cluster ions as they pass through the lower atmosphere. Neutral molecular clusters such as dimers and complexes are expected to make a small contribution to the radiative balance, but atmospheric absorption by charged clusters has not hitherto been observed. In an atmospheric experiment, a thermopile filter radiometer tuned to a 9.15μm absorption band, associated with infra-red absorption of molecular cluster ions, was used to monitor changes following events identified by a cosmic ray telescope sensitive to high energy (>400MeV) particles, principally muons. The change in longwave radiation in this absorption band due to molecular cluster ions is 7 mWm$^{-2}$. The integrated atmospheric energy change for each event is 2Jm$^{-2}$, representing an amplification factor of $10^{12}$ compared to the estimated energy density of a typical air shower. This absorption is expected to occur continuously and globally, but calculations suggest that it has only a small effect on climate.

## 1. Introduction

Atmospheric molecular cluster ions (MCI) are bipolar charged species formed by ionisation from cosmic rays. In the troposphere this ionisation is mainly from a cascade of secondary subatomic particles, such as muons and electrons, produced by the decay of energetic primary cosmic rays, and natural radioactivity emitted from the surface [1]. The cascade of secondary particles from a single primary cosmic ray is known as an "air shower" [2]. MCI are generated when core positive ions e.g. $N_2^+$, or electrons attached to electrophilic molecules (e.g. $O_2^-$), rapidly cluster with polar ligands that are hydrogen bonded to the core ion e.g. $HSO_4^-(H_2SO_4)_m(H_2O)_n$ or $H_3O^+(H_2O)_n$ [3,4]. The wide range of hydrogen-bonded atmospheric MCI species absorb and emit infra-red (IR) radiation, for example IR transitions associated with bond stretching and bending for the gas phase protonated water dimer $H_3O^+(H_2O)_2$ have been measured in the laboratory [5]. Although the contributions of neutral molecular clusters, such as the water oligomer $(H_2O)_n$ and hydrated complexes (e.g. $O_2$-$H_2O$) to atmospheric radiative transfer via IR absorption are being actively investigated [6,7], the radiative properties of atmospheric MCI – which are a direct effect on the atmosphere's radiation budget - have not hitherto been considered. Responses of MCI to sudden decreases in cosmic rays have already been indirectly demonstrated through atmospheric electricity changes [4], but here we present evidence that MCI formed in the atmosphere by cosmic rays also absorb IR radiation within the broad absorption band previously identified in laboratory experiments.

Spectroscopic measurements in the laboratory with artificially generated MCI detected IR absorption of 1-3% in two bands centred on 9.15 and 12.3 μm [8,9], with MCI columnar concentrations of $10^{13}$ m$^{-2}$. As the atmospheric MCI columnar concentration is estimated to be $10^{14}$ m$^{-2}$, detectable absorption is therefore expected. Based on these laboratory data, the experiment described in this paper was devised to search for the effects of atmospheric MCI on longwave radiation. In this paper we report the atmospheric response of a narrowband thermopile radiometer tuned to the 9.15 μm MCI absorption band, following cosmic ray events ionising the atmospheric column above the radiometer.

## 2. Experiment

The sensor used in this experiment was an atmospheric thermopile radiometer, with a filter spectrally tuned to pass radiation in the region centred upon 9.15 μm with a bandwidth (FWHM) of 0.9 μm (i.e ±5% of band centre) [10], with a stable low noise amplifier [11] for signal conditioning. A small cosmic ray telescope using vertically stacked Geiger counters was located close to the filter radiometer, to detect the high-energy particles [12] creating atmospheric MCI over the radiometer. Adjacent broadband thermopile radiometers were used to monitor downwelling atmospheric short wave (SW, 0.3-3 μm) radiation, emitted by the Sun, and downwelling, terrestrially emitted, long wave (LW, 4.5-42 μm) radiation. The cosmic ray telescope indicates an "event" when both its detectors are triggered by high-energy particles travelling down through the atmosphere (the false triggering rate has been shown to be negligible [12]). Subsidiary experiments were carried out to investigate the energy sensitivity of the detector by placing it beneath varying quantities of lead and concrete, selected to absorb different energies of particle [e.g. 13]. These showed that the telescope responds to particles with energy >400 MeV, which, at the surface, are almost all muons (mean energy 2GeV) [14]. Previous experiments with the same apparatus showed, firstly, the electrical conductivity of the air, which is approximately proportional to the atmospheric MCI concentration, increased after muon events [15]. Secondly, the filter radiometer employed here has already been shown to respond to changes in atmospheric MCI in a calibration experiment using direct measurements of MCI [16].

The cosmic ray telescope was housed in a waterproof enclosure at a semi-rural UK site (51.8929N, 2.1300W). The filter radiometer was installed above it on the roof of a nearby log cabin building, adjacent to the broadband radiometers of a Kipp and Zonen CNR1 instrument (Figure 1). A Campbell CR3000X data logger was used to count the cosmic ray telescope events and to log the radiometer data. Radiometer values were sampled every 20s, which is the timescale for the radiometer's thermopile sensor to fully respond to a step change. This slow instantaneous sampling also circumvents the possibility of crosstalk between the instruments, since there is up to 20s delay between the triggering signal and recording of the radiometer response. This means, for example, that we can reject the possibility of the radiometer signal conditioning electronics itself suffering direct ionisation from the energetic particle event, as the fast response (<<1s) of the amplifier [11] will allow induced charge to dissipate rapidly before the next sample is taken. (No such effect is likely in the radiometer's sensing thermopile, as it is not a semiconductor device.) The 20s data samples were only saved for 400s either side of a triggering event, and 5 minute averages were also recorded. The experiment ran from July 2008 to June 2009, over which the mean high-energy particle flux was 33 m$^{-2}$ster$^{-1}$s$^{-1}$, consistent with the mean flux of >1GeV muons expected at the surface [14].

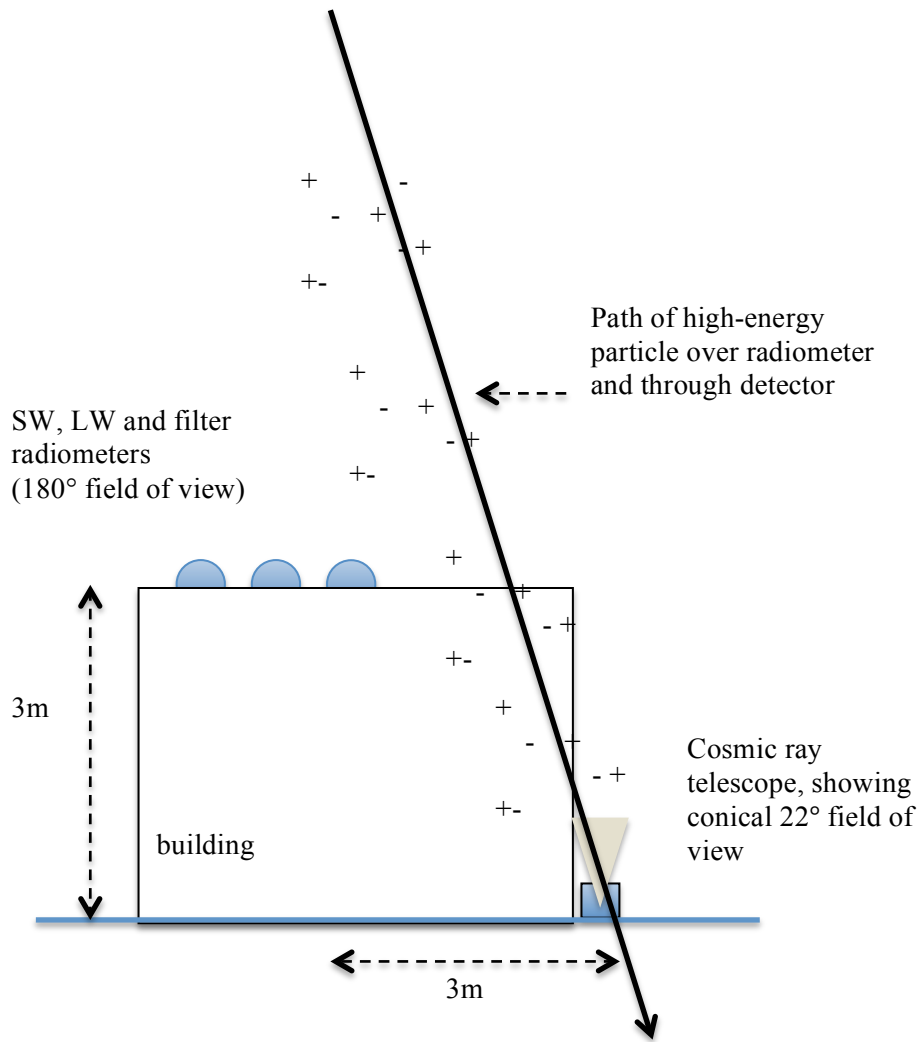

Figure 1 Schematic arrangement of the experiment, (not to scale) showing the radiometers mounted on a building and the adjacent cosmic ray telescope. The telescope can detect particles in a cone 11° from the vertical, with the geometry implying that approximately half the particles detected have passed over the radiometer at a height of >15m.

## 3. Results

The IR radiation measured by the filter radiometer is calibrated with reference to the blackbody atmospheric brightness temperature in its passband, calculated with data from the CNR1 LW radiometer [17]. A positive response from the filter radiometer indicates emission in the MCI wavelength range in the column above the radiometer, and a negative signal indicates absorption with respect to the blackbody background. Figure 2 shows a time series of the combined data from all instruments for approximately six months in 2009. The typical variations in each quantity can be seen, particularly the range in LW down, which is greatest in warm, cloudy skies and least in the nocturnal clear sky. The filter radiometer output shows a diurnal variation, with more absorption in the band at night and more emission during the day.

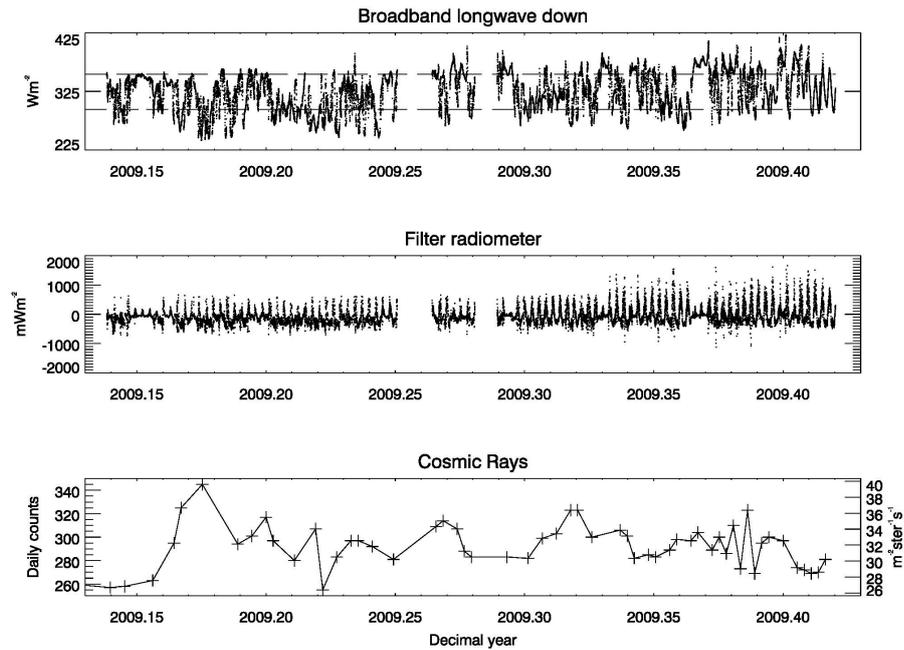

Figure 2 Time series of typical data, from 16th February – 6th June 2009 showing (upper panel) broadband longwave down (5 minute averages) with the upper (354 Wm$^{-2}$) and lower (295 Wm$^{-2}$) quartiles as horizontal dashed lines. The middle panel shows the filter radiometer signal (5 minute averages), and the lower panel presents daily averages of the raw muon flux, not temperature or pressure corrected, as both daily counts (left hand axis) and flux (right hand axis).

The data analysis approach taken is to separate out the response around each cosmic ray telescope event, which are averaged together ("composited") to obtain the typical response. Compositing many events (also referred to as a "superposed epoch" or a "Chree" analysis) is a well-established technique for extracting signals despite background variability when there is thought to be a triggering event in an independent dataset [18]. Although the atmospheric pressure and temperature do affect the surface muon flux [19,20] plotted here as raw data in figure 2 (lower panel), our compositing approach means that we only analyse the immediate change in the local radiative response to air showers (detected by a >400MeV particle entering the cosmic ray telescope) on timescales much shorter than the pressure and temperature changes (which would average out in the analysis in any case).

### 3.1 Composited data

Figure 3 shows infra-red changes measured by the filter radiometer from 25th July 2008 to 2nd June 2009, plotted as a composite around high-energy particle events, occurring at time $t = 0$ in each case. All the available data has been used to form the composite, with each event normalised by subtracting the median IR absorption over the 400s preceding the event. Data from 400s before each event is not significantly different to the background data at 800s before each event. In the upper panel, a difference in the median response after the triggering event can be seen to emerge from the variability. The background variability was calculated from multiple realisations of randomised data chosen from the no event period (that before each event), using the same number of random values as the number of real data points for the sample concerned (shown in the lower panel). The median response to all events shows a statistically significant absorption of up to 4 mWm$^{-2}$, which then recovers to the background level (this is not statistically different from the pre-event background). In contrast, no effect is seen if a similar analysis is performed on the broadband downwelling LW measured with the CNR1 net radiometer.

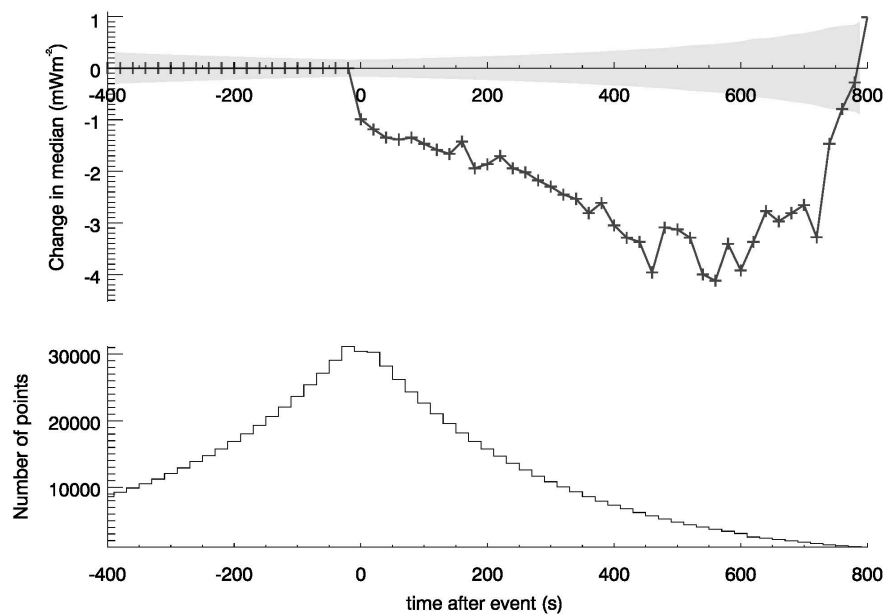

Figure 3 Infra-red filter radiometer data (from 25th July 2008 to 2nd June 2009), around high energy particle events triggering the counter, considered to occur at 0s. The response to 31398 events is shown, with the median during the 400s before each event subtracted in each case. The upper panel indicates the change in median filter radiometer signal following each event (grey line). The shaded region indicates the natural variability expected for the number of points in the composite, determined from periods during which no high-energy particle events were recorded. The pre-event variability is calculated from the usual confidence range on the median for a non-Gaussian distribution (1.58 x the interquartile range divided by the square root of the number of points [21]). The lower panel indicates the number of points contributing at each time. By 800s after the event there are few (~100) data points, as there is a high probability that the next event has occurred by then. If another event occurs within 800s, only data until just before the next event is included.

As the IR radiation emitted by water vapour in cloud will contribute to the filter radiometer signal, one expectation might be that the direct IR absorption of MCI is most apparent in clear sky. In cloudy sky, the additional absorption from water vapour in the clouds would contribute, and could obscure the absorption effect. Situations in which clouds are largely absent are chosen by selecting the lower quartile of the LW measurements. The lower quartile (LW <295 Wm$^{-2}$) is dominated by measurements on clear winter nights, whereas the upper quartile (LW >354 Wm$^{-2}$) occurs mainly under cloudy conditions in the spring and summer.

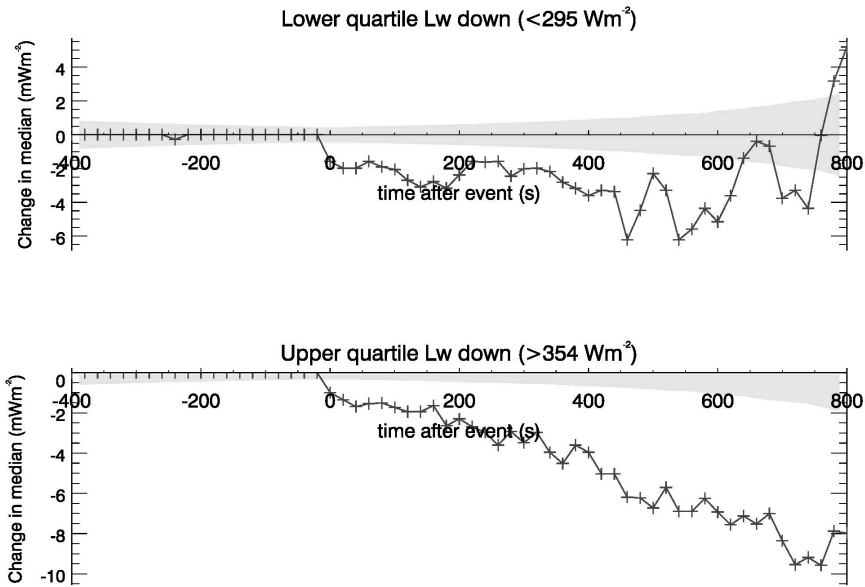

Figure 4 Change in median filter radiometer output (similar to the top panel of figure 3), sorted by downwelling longwave radiation. The upper panel shows the lower quartile (<295 Wm$^{-2}$), 6295 events, and the lower panel the upper quartile (>354 Wm$^{-2}$), 8164 events. The shaded region indicates the natural variability expected for the number of points in the composite, determined from periods during which no high-energy particle events were recorded (as for figure 3).

The two situations are compared in Figure 4. This indicates that the same sign of response is apparent in both cloudy and clear skies, with the absorption appearing both enhanced, and of longer duration in cloudy skies (Figure 4, lower panel) compared to clear, cold skies (Figure 4, upper panel). The apparent enhancement and lack of recovery of the effect in cloudy sky is likely to be related to water vapour absorption in the passband.

**3.2 Timescale of the effect**

If the absorption seen is a response to ionisation above the radiometer caused by cosmic rays, then the shape of the response should reflect the physics of atmospheric MCI. The observations are consistent with increased ionisation after the events, followed by a recovery as the MCI are lost by attachment to atmospheric aerosol particles or self-recombination.

The removal of MCI by self-recombination will be considered first, which are hypothesised to cause the recovery back to pre-event conditions from the IR minimum shown in Figure 3. In relatively clean air, recombination of oppositely charged MCI dominates and the MCI lifetime $t_r$ is given by

$$t_r = \frac{1}{\alpha n} \quad (1)$$

where $\alpha$ is the recombination coefficient (1.6 x 10$^{-6}$ cm$^3$s$^{-1}$ for typical surface conditions [3]) and $n$ is the MCI concentration. Using the differential of a spline fitted to the data to identify the minimum gives a recovery period of (280±60)s. Using (1) to estimate the MCI concentration from the recovery time gives $n$= (2300±500) cm$^{-3}$, which is consistent with slightly enhanced MCI concentrations over typical measured background levels [3], as expected from a burst of MCI created by the cascade associated with the high-energy particle.

The initial slow linear increase in absorption following the cosmic ray event trigger occurs over (520±60) s. As the high-energy particles are relativistic, and initial ionisation occurs over nanoseconds, this will be related to the spread of the MCI in the atmosphere above the radiometer. From the experiment geometry (Figure 1), any particle triggering the telescope must pass directly over the radiometer at a height of at least 15m, and will form MCI only around its track, as will most of the other secondary particles in the cascade, which do not deviate far in direction of motion from that of the primary. The absorption effect seen can be explained by fresh MCI drifting towards the radiometer after a burst of ionisation. The IR absorption then slowly returns to pre-event levels as the MCI are lost by recombination. Mobility $\mu$, defined in equation (2), can be used to determine the drift speed $v$ attained by charged clusters in an electric field of magnitude $E$ as

$$v = \mu E \quad (4).$$

As the typical atmospheric electric field is 100 Vm$^{-1}$, and is downwards directed, then the positive MCI formed will drift downwards, and the negative MCI upwards at 1 cms$^{-1}$, moving ~5m in 500s. In comparison, molecular diffusion would be negligible in the time considered, and advection by the wind would distribute the MCI over at least 100m.

4. Discussion

4.1 Sources of variability

The cosmic ray telescope detects high-energy particles in a solid angle of 0.34 steradians from the vertical, whereas the radiometer responds to changes over almost a hemisphere (2π steradians). The radiometer is therefore affected by IR radiation over a solid angle twenty times greater than the muon detector, responding to IR changes from clouds and water vapour as well as the atmospheric IR absorption from MCI reported here for the first time. This will cause background variability in the radiometer signal unrelated to ionisation, and may explain why many events need to be composited to see an effect. Muons and the ionisation they produce in typical cascades of secondary particles from one primary cosmic ray particle (an "air shower") [2] will be insufficient to entirely account for our findings [22]. However, muons are not the only relevant ionising radiation produced in an air shower [23], so the events detected are therefore likely to indicate ionisation well above the radiometer from other particles created in the same air shower.

The contribution from ionisation at different altitudes to the measured IR response results from both the number of ions created by a primary particle, and the IR radiation emitted in the atmosphere. The IR radiation received in the instrument's passband originates from atmospheric LW emission, decreasing with temperature from the surface. Primaries >10GeV are most likely to create muons at the rate we detect, and the ionisation yield function, which varies with altitude and primary energy, shows that ionisation from 10-100GeV primary protons peaks at an altitude of ~15km, and is an order of magnitude less at the surface [23]. Over the same height, a temperature change of about 50K reduces the emitted radiation in the passband by a factor of 5. Hence, combining the two effects, it is apparent that the ions created at 10-15km will provide the dominant contribution to the radiometer signal. This altitude is also consistent with our measurements of the MCI absorption signal in clear and cloudy sky, showing that the characteristic timescales associated with MCI are not apparent in cloudy conditions (figure 4). This is what would be expected from clouds occurring beneath the ion absorption region at 10 to 15km, and their IR emission hiding the ion absorption effect above.

Variability in the radiative response will occur from MCI formed in the radiometer field of view by particles not in the acceptance cone of the telescope, and from some cosmic ray secondaries triggering the telescope without passing over the radiometer. The minimum detectable size of air shower would in principle create particles only over the area defined by the distance between the radiometer and cosmic ray telescope (~5 m), but these low-energy showers are unlikely to create energetic enough secondaries to trigger our telescope. Higher energy (GeV) primaries generate larger air showers, of at least 40m lateral extent [2], containing muons that can trigger our detector. The magnitude of the signal would be related to how close the instruments were to the core of the shower where most ions are made, creating additional variability in our response. This could be readily confirmed by experiments with separated cosmic ray detectors, which are often co-located with detailed meteorological measurements [e.g. 20,24]. Additional measurements could also help understand how much ionisation is associated with each triggering event, which would permit calculation of the IR radiation absorbed per ion created.

In [23], the modelled total ionisation yield as a function of altitude for three different primary cosmic ray energies of primary cosmic ray, 200 MeV, 1GeV and 100 GeV, were calculated, including the separate contributions from the three principal components of the cascade: electromagnetic, muon and hadronic. For the 200MeV primary, the only important ionising contribution is from the hadronic component, however, this energy of primary is not relevant to our study as it is almost certainly below the threshold for the cosmic ray telescope. For 1GeV primaries, the ionisation is dominated by the electromagnetic component at the highest altitudes, hadrons at middle altitudes and muons very close to the ground. For the high energy (100GeV) primaries, the hadronic component is low, with the secondary muons dominating ionisation in the lower troposphere and the electromagnetic component at higher altitudes. It is worth noting here that the radiative effect of any greenhouse gas varies with altitude, as it re-radiates less, and so traps more, at higher altitudes where the temperature is lower. This will also be true for the radiative effect of MCI. Therefore the radiative effect we detect is quite likely to be associated with the electromagnetic component of the cascade that generates ionisation outside the boundary layer.

**4.2 Radiative contribution**

In estimating the potential radiative forcing associated with the IR absorption described above, it is important to bear in mind that the effect seen has only been measured in the radiometer's passband and could also be occurring in other spectral regions, e.g the 12.3μm band seen in the laboratory [8]. Thus the radiative effect evaluated here is expected to be an underestimate.

The events used to form the composites shown in figures 3 and 4 occurred at an average rate of 12 hr$^{-1}$ and the additional IR absorption averaged ~2.5 mWm$^{-2}$ over 800 s. Therefore the average power trapped is 2.5×12×800/(60×60) ~7 mWm$^{-2}$. Ionisation chamber data indicates that the solar cycle variation in atmospheric ion production rate is up to 15% [1]. Thus we expect the solar modulation of cosmic rays to induce a variation of up to 15% of this radiative effect of 7 mWm$^{-2}$ i.e. ~1 mWm$^{-2}$. Reconstructions of the long-term variation in cosmic ray fluxes give centennial-scale variations on of the same order as the typical solar cycle changes discussed above [25, 26] and hence we expect changes on centennial timescales also to be only about 1mWm$^{-2}$.

The radiative climate forcing of MCI-induced absorption would be the reduction in upward-going IR radiation at the top of the troposphere. This will be of the same order of magnitude as the change in downward IR re-radiated by the atmosphere, which is what we have detected here. Full radiative transfer calculations will be needed to evaluate the top of atmosphere radiative forcing from the observed change in downward radiation. A radiative forcing of 1mWm$^{-2}$ is small compared to other known factors: for example the change in trace greenhouse gas concentrations over the past century gives about 2.5 Wm$^{-2}$ and the estimated change in total solar irradiance gives a radiative forcing of about 0.2 Wm$^{-2}$ [27]. Our results, which represent the first quantification of this effect in the atmosphere, allow us to conclude that this is unlikely to be a significant factor modulating Earth's energy balance, although it is expected to occur both globally and continuously.

Each detected event generates an integrated energy transfer of 1.9 Jm$^{-2}$, whereas a typical air shower of 40m radius, generated by a 10GeV primary [2,23], gives an incoming mean energy density of 2 MeVm$^{-2}$ ($10^{-13}$ Jm$^{-2}$). Our mechanism therefore represents a substantial energy amplification, of $10^{12}$, which is direct, in comparison to other proposed mechanisms for radiative effects of cosmic rays [28]. Finally, an interesting point arises from these results in terms of monitoring cloud cover from space. The attenuation of the outgoing longwave radiation is used in retrieval algorithms to determine cloud cover in remote sensing data. Hence, using the passband of our experiment to derive low cloud from satellite data [29 and references therein] could contribute to a solar imprint in observations of satellite-derived global cloud cover.

**Acknowledgements**

This work was partially funded by the UK Science and Technology Facilities Council. We thank Prof R G Harrison (University of Reading) and Prof T Sloan (Lancaster University) for helpful comments.

**References**

[1] G A Bazilevskaya et al, Cosmic ray induced ion production in the atmosphere, *Space Science Reviews*, 137, 149-173 doi: 10.1007/s11214-008-9339-y (2008)
[2] K Griesen, Cosmic ray air showers, *Ann. Rev. Nuc. Sci.*, **10**, 63-108 (1960)
[3] R G Harrison and K S Carslaw, Ion-aerosol-cloud processes in the lower atmosphere *Reviews of Geophysics* **41** (3), 1012, (2003)
[4] W E Cobb, Evidence of a solar influence on the atmospheric electric elements at Mauna Loa Observatory, *Mon. Wea. Rev.*, **95**, 12, 905 (1967)
[5] K Asmis et al, Gas-phase infrared spectrum of the protonated water dimer, *Science,* 299, 1375-1377 (2003)
[6] W Klemperer and V Vaida, Molecular complexes in close and far away, *PNAS* 1-3, 28, 10584-10588 (2006)
[7] K P Shine et al, The water vapour continuum: brief history and recent developments, *Surveys in Geophysics* doi 10.1007/s10712-011-9170-y (2012)
[8] H R Carlon, Infrared absorption and ion content of moist atmospheric air, *Infrared Physics* 22, 43-49 (1982)
[9] K L Aplin and R A McPheat, Absorption of infra-red radiation by atmospheric molecular cluster-ions, *J. Atmos. Sol-Terr. Phys*, 67, 775-783 (2005)
[10] K L Aplin and R A McPheat, An infra-red filter radiometer for atmospheric cluster-ion detection, *Rev. Sci. Instrum* 79, 106107 (2008)
[11] R G Harrison and J R Knight, Thermopile radiometer signal conditioning for surface atmospheric radiation measurements *Rev Sci Instrum* **77**, 116105 (2006)
[12] K L Aplin and R G Harrison, Compact cosmic ray detector for unattended atmospheric ionization monitoring, *Rev Sci Instrum*, **81**, 124501 (2010)
[13] W B Gilboy et al, Muon radiography of large industrial structures, *Nuc. Inst. Meth. Phys. Res. B.* **263**, 317-319 (2007)
[14] K Nakamura et al (Particle Data Group) 2011 Review of particle physics, *J. Phys G.* 37, 075021 (2011)
[15] R G Harrison and K L Aplin, Atmospheric condensation nuclei formation and high-energy radiation *J. Atmos. Solar-Terrestrial Physics* **63**, 17, 1811-1819 (2001)
[16] K L Aplin, Composition and measurement of charged atmospheric clusters *Space Sci Revs* **137**, 1-4, 213-224 (2008)
[17] M J Rycroft et al, Global electric circuit coupling between the space environment and the troposphere, *J. Atmos. Sol-Terr. Phys.*, doi: 10.1016/j.jastp.2012.03.015 (2012)
[18] S E Forbush et al, Statistical procedures for test Chree analysis results, *Proc 17$^{th}$ Int. Cosmic Ray Conf.*, Paris, France, 4 pp 47-51 (1981)
[19] A M Hillas, *Cosmic rays*, Pergamon Press, Oxford (1972)
[20] P Adamson et al (MINOS Collaboration), Observations of muon intensity variation by season with the MINOS far detector, *Phys. Rev. D.,* **81**, 012001 (2010)

[21] R McGill et al, Variations of box plots. *The American Statistician* **32**, 12–16 (1978)
[22] T Sloan, private communication (2012)
[23] I G Usoskin and G Kovaltsov, Cosmic ray induced ionization in the atmosphere: Full modeling and practical applications, J. Geophys. Res., 111, D21206 (2006)
[24] B Keilhauer et al, Impact of varying atmospheric profiles on extensive air shower observation: atmospheric density and primary mass reconstruction, *Astropart. Phys*. **22**, 249 (2004)
[25] I G Usoskin et al., A physical reconstruction of cosmic ray intensity since 1610, J. Geophys. Res., 107(A11), 1374, doi:10.1029/2002JA009343 (2002)
[26] F Steinhilber et al, Solar modulation during the Holocene, *Astrophys. Space Sci. Trans*., **4,** 1–6, doi:10.5194/astra-4-1-008 (2008)
[27] M Lockwood, Solar change and climate: an update in the light of the current exceptional solar minimum, *Proc. R. Soc. A*, **466** 303-329, doi:10.1098/rspa.2009.0519 (2010)
[28] B A Tinsley and G W Deen, Apparent tropospheric response to MeV-GeV particle flux variations: a connection via electrofreezing of supercooled water in high-level clouds? *J. Geophys. Res*., **96**, D12, 22,283-22,296, doi:10.1029/91JD02473 (1991)
[29] N D Marsh and H Svensmark, Low cloud properties influenced by cosmic rays, *Phys. Rev. Lett*. 85, 23, 5004-500 (2000)